%% file: main.tex
\documentclass[conference]{IEEEtran}

 \usepackage{mathtools}
\usepackage{amssymb}
\usepackage{enumerate}

\usepackage[normalem]{ulem}
\usepackage{soul} 

\usepackage{balance} 
\usepackage{booktabs} 
\usepackage{multirow}
\usepackage{subcaption}
\usepackage{algorithm}
\usepackage{algpseudocode}
\usepackage[greek,english]{babel}
\usepackage{lipsum}
\usepackage{mathtools}
\usepackage{cuted}
\usepackage{gensymb}
\usepackage{amsmath} 
\usepackage{booktabs}
\usepackage[justification=centering]{caption}
\usepackage{color}
 \usepackage{lipsum}  
\usepackage{hyperref} 
\usepackage{tikz}

\usepackage{enumitem}

\begin{document}

\title{Learning to Spend: Model Predictive Control for Budgeting under Non-Stationary Returns}

\newcommand{\cs}[1]{{\bf \color{red}{cs: #1}}}
 
\newcounter{regime}
\renewcommand{\theregime}{R\arabic{regime}}

\newcommand{\regimepara}[1]{%
  \refstepcounter{regime}%
  \paragraph*{\textbf{\theregime: #1}}%
}

 \author{ \IEEEauthorblockN{Nilavra Pathak\IEEEauthorrefmark{1},   Smriti Shyamal\IEEEauthorrefmark{1}, Prasant Mhasker\IEEEauthorrefmark{2}, Christopher Swartz  \IEEEauthorrefmark{2}}   
 \IEEEauthorblockA{\IEEEauthorrefmark{1}Marketing Data Science, Expedia Group}   
 \IEEEauthorblockA{\IEEEauthorrefmark{2} Department of Chemical Engineering, McMaster University}

 \IEEEauthorblockA{npathak@expediagroup.com\IEEEauthorrefmark{1}, sshyamal@expediagroup.com\IEEEauthorrefmark{1},  mhaskar@mcmaster.ca\IEEEauthorrefmark{2}, swartzc@mcmaster.ca\IEEEauthorrefmark{2}}}

  \maketitle
\begin{abstract}
We study finite-horizon budget allocation as a closed-loop economic control problem and
evaluate receding-horizon Model Predictive Control (MPC) relative to reactive budgeting
policies. Budgets are allocated periodically under execution noise and operational
constraints, while return efficiency may evolve over time. Using a controlled simulation
framework motivated by digital marketing, we compare reactive pacing to MPC across
environments with increasing degrees of non-stationarity.
Our results show that non-stationarity alone does not justify predictive control.
When return dynamics are stationary or evolve through unpredictable stochastic drift,
MPC offers no systematic advantage over reactive baselines. By contrast, when return
efficiency exhibits predictable structure over the planning horizon, that is captured through an underlying model, MPC consistently
outperforms reactive budgeting by exploiting intertemporal trade-offs. 
\end{abstract}


 \input{Introduction}
  \input{Problem.tex}
 \input{Data_Generation}

 \input{Policies}
 \input{Experimental_Setup}

 \input{Experimental_Results}
   \input{conclusion}

 \bibliographystyle{unsrt}

 \bibliography{bib.bib} 

 \input{Appendix}

 \end{document}

%% file: Introduction.tex
\section{Introduction}

Marketing budgets are allocated periodically under uncertainty, execution noise, and operational constraints. Planned spend does not translate deterministically into realized exposure or conversions due to auction dynamics, pacing systems, and competitive effects. At the same time, the effectiveness of spend may evolve over time as a result of market conditions, saturation, or platform behavior. These features make budget allocation a closed-loop decision problem with delayed and noisy feedback. Solving these problems is critical for large companies that manage multi-billion-dollar marketing budgets; for example, Expedia Group increased its annual marketing investment to approximately \$6.8 billion in 2024, and marketing expense represents one of the largest operating cost categories for online travel agents. Such scale underscores the practical and economic importance of principled budget allocation and pacing strategies in commercial deployments~\cite{expedia10k,expedia10q}.

A large literature focuses on estimating spend--return relationships from historical data,
including marketing mix models and related econometric approaches
\cite{hanssens2003market, DekimpeHanssens1995Persistence}. These methods provide interpretable
response curves and remain widely used in practice. However, they are fundamentally
descriptive: response functions are estimated offline and treated as fixed inputs to
downstream planning, despite the underlying environment being non-stationary and affected
by prior allocation decisions.
Building on these response estimation models, budget allocation decisions are typically performed by constrained
optimization, either statically or through periodic re-solving
\cite{hatano2015lagrangian, luzon2022dynamic, zhao2019unified}. While effective for short-term
or marketing channel-level decisions, these approaches generally assume stable response behavior
over the planning horizon and do not explicitly account for closed-loop adaptation under
execution noise.

Sequential decision-making frameworks based on reinforcement learning and bandits relax
these assumptions by optimizing long-run performance
\cite{archak2012budget, cai2023kdd_autobid, yan2023endtoend}. However, such methods often
require large volumes of logged data, rely on implicit policies, and involve exploration
risk that is difficult to justify in high-stakes, low-frequency budgeting settings such as
quarterly planning. A complementary perspective arises from control theory. Classical models such as
Nerlove--Arrow and Vidale--Wolfe explicitly model advertising as a dynamical system
\cite{nerlove1962optimal, sethi1973optimal}. While analytically appealing, their open-loop
solutions are highly sensitive to model assumptions and do not naturally incorporate
operational constraints, execution noise, or repeated re-planning.

Model Predictive Control (MPC) addresses these limitations by repeatedly solving a
finite-horizon optimization problem using updated system information while respecting
constraints \cite{morari1999model, rawlings2017model}. MPC has proven effective in domains
characterized by delayed feedback, uncertainty, and feasibility constraints, making it a
natural candidate for budget allocation problems. Despite this alignment, it remains unclear when predictive control meaningfully
outperforms simpler reactive budgeting heuristics. Non-stationarity alone does not
guarantee gains from receding-horizon optimization: when changes in return efficiency are
irregular or difficult to forecast, predictive control may offer little advantage.
Conversely, when non-stationarity exhibits predictable structure over the planning horizon,
anticipatory allocation may unlock substantial improvements~\cite{li2019online, yu2020power}.

In this paper, we study budget allocation through this control-theoretic lens. Using
marketing budget planning as a motivating application, we compare reactive pacing strategies against
receding-horizon MPC under execution noise and budget constraints, with the goal of
characterizing when predictive control improves closed-loop allocation outcomes. \textbf{Core contributions of the paper are:}
\begin{itemize}
    \item We formulate finite-horizon budget allocation as a constrained closed-loop
    control problem and apply receding-horizon MPC to periodic budgeting decisions.

    \item We characterize the performance of predictive control under different forms
    of non-stationarity, distinguishing unpredictable drift from structured, forecastable
    variation.

    \item Simulation experiments demonstrate that MPC provides value only when
    non-stationarity is predictable over the planning horizon, and offers no advantage
    otherwise.
\end{itemize}

%% file: Problem.tex
\section{Problem Setup}
\label{sec:problem}

We study finite-horizon budgeting as a closed-loop decision problem in which a planner
sets weekly spend targets and a downstream execution layer converts those targets into
realized spend under uncertainty, reflecting modern digital advertising delivery systems
\cite{zhao2019unified, chen2011real, cai2017realtime, agarwal2014budgetpacing, balseiro2024fieldguide}.
Time is discretized in weeks over an evaluation quarter $t=1,\dots,W$ (with $W=12$). At
the start of week $t$, the planner selects a planned weekly budget $b_t\ge 0$ given the
remaining budget $B_t^{\mathrm{rem}}$ (with $B_1^{\mathrm{rem}}=B_Q$ and
$B_{t+1}^{\mathrm{rem}}=B_t^{\mathrm{rem}}-s_t$). Realized spend is a noisy actuation of
the plan,
\begin{equation}
s_t = \sigma_t\!\left(b_t,\Omega_t\right) + \omega_t,
\label{eq:exec_map}
\end{equation}
where $\Omega_t$ captures exogenous market conditions and $\omega_t$ is execution noise.
Weekly return is generated from realized spend via a nonlinear, saturating response with
latent efficiency parameters that may vary over time,
\begin{equation}
r_t = \rho\!\left(s_t;\theta_t\right) + \varepsilon_t,
\label{eq:return_map}
\end{equation}
where $\varepsilon_t$ is observation noise. The planner solves the finite-horizon
stochastic program
\begin{equation}
\max_{b_{1:W}} \ \mathbb{E}\!\left[\sum_{t=1}^{W} r_t\right]
\quad \text{s.t.} \quad
\sum_{t=1}^{W} b_t \le B_Q,\ \ b_t\ge 0,
\label{eq:objective}
\end{equation}
optionally with additional operational guardrails such as bounded week-to-week budget
changes to enforce stable spend plans.

%% file: Data_Generation.tex
\section{Data Generation Process}
\label{sec:data_generation}

We construct a synthetic marketing environment that generates both historical and
evaluation data independently of any control policy. The environment specifies
(i) how quarterly budgets are constructed, (ii) how planned budgets translate into
realized spend through a noisy execution layer, and (iii) how realized spend produces
returns via a latent nonlinear response that may evolve over time. This separation
ensures that all policies are evaluated on identical underlying system dynamics and
that performance differences arise from control logic rather than simulator bias.

\subsection{Timeline and interaction loop}
\label{sec:dg_timeline}

Time is measured in days, while control decisions are made at a weekly cadence. Each
trial consists of two contiguous segments
\begin{itemize}
    \item \textbf{Historical period}: $Y$ years of daily spend and return observations,
    used to initialize policies (e.g., fitting initial spend--return
     predictors for MPC). No data from the evaluation period is used for initialization.
    This mirrors common practice in marketing systems, where historical logs are used to
    initialize planning models before budgets are executed online.

    \item \textbf{Evaluation period}: one held-out quarter of $W=12$ weeks, during which
    policies interact with the environment in closed loop
\end{itemize}

Weeks are indexed by $t \in \{1,\dots,W\}$ and days within a week by
$d \in \{1,\dots,7\}$. At the start of week $t$, a policy selects a planned weekly budget
$b_t$. The environment generates realized daily spend
$\{s_{t,d}\}_{d=1}^7$ and daily returns $\{r_{t,d}\}_{d=1}^7$, which are aggregated to
weekly totals
\[
s_t = \sum_{d=1}^7 s_{t,d},
\qquad
r_t = \sum_{d=1}^7 r_{t,d}.
\]
Policies observe $(s_t, r_t)$, update internal state, and proceed to week $t+1$. Within a
trial, all policies face identical budgets, constraints, and stochastic realizations
(paired evaluation).

\subsection{Budget construction}
\label{sec:budget_construction}

Let $B^{(q)}$ denote the evaluation-quarter budget. To ensure realistic budget magnitudes
without leaking future information, $B^{(q)}$ is constructed using only the final year of
the historical period. Specifically, annual spend is computed from historical data,
perturbed by a stochastic year-over-year scaling factor, and allocated across months using
historical monthly proportions; the quarterly budget is defined as the sum of the
corresponding three months. Full construction details are provided in
Appendix~\ref{app:budget_construction}.

\subsection{Execution layer: noisy spend tracking}
\label{sec:spend_tracking}

In real advertising platforms, realized spend deviates from planned budgets due to
auction outcomes, inventory constraints, and pacing logic
\cite{chen2011real, cai2017realtime, zhao2019unified}. We model this via a noisy tracking
process that maps weekly budgets to daily spend. Let $\bar{s}_t = b_t/7$ denote the
nominal daily target. Daily spend evolves as
\begin{equation}
s_{t,d}
=
s_{t,d-1}
+ \alpha_t\big(\bar{s}_t - s_{t,d-1}\big)
+ \xi_{t,d},
\qquad d = 1,\dots,7,
\label{eq:spend_tracking_exp}
\end{equation}
where $\alpha_t \in (0,1)$ is an adaptation (tracking) rate and $\xi_{t,d}$ is mean-zero
delivery noise with variance proportional to the target level. This execution layer
induces realistic friction: even identical plans $b_t$ yield stochastic realized spend
$s_t$, which affects both reactive and predictive controllers.

\subsection{Return layer: latent nonlinear response}
\label{sec:return_generation}

Returns are generated from realized spend via a nonlinear, saturating
\emph{Richards-type response curve}~\cite{richards1959flexible}, which generalizes
exponential saturation by allowing asymmetric curvature and variable saturation rates.
This choice reflects empirical evidence that advertising response curves deviate
systematically from simple exponential forms \cite{Wang2023}. Daily returns satisfy
\begin{equation}
r_{t,d}
=
\rho_{\mathrm{Richards}}\big(s_{t,d}; \theta_t\big)
+ \varepsilon_{t,d},
\label{eq:return_model_exp}
\end{equation}
where $\theta_t$ are time-varying parameters governing scale and elasticity, and
$\varepsilon_{t,d}$ captures observation noise.

For control, MPC deliberately assumes a simpler exponential-saturation response model.
This intentional functional misspecification isolates the role of
\emph{predictability in latent efficiency} rather than estimator fidelity: any
performance gains must arise from forecasting structure in $\theta_t$, not from matching
the true response shape. Unless otherwise stated, the same environment class generates
both historical and evaluation data.

\subsection{Operating regimes}
\label{sec:regimes}

We study three \emph{operating regimes} that differ in the predictability of return
efficiency over the planning horizon
\begin{enumerate}[=\textbf{R\arabic*:}]
    \item \textbf{Static (stationary) regime.}
    Parameters are fixed during evaluation, with $\theta_t = \theta$ and
    $\alpha_t = \alpha$. Outcomes remain noisy, but the underlying response and execution
    dynamics do not drift.

    \item \textbf{Random-walk drift regime.}
    Return efficiency evolves smoothly but stochastically
    \begin{equation}
    \log \theta_{t+1}
    =
    \log \theta_t + \eta_t,
    \qquad
    \eta_t \sim \mathcal{N}(0, \sigma_\eta^2),
    \label{eq:rw_theta}
    \end{equation}
    where $\eta_t$ is the random-walk innovation. Execution dynamics may also drift
    analogously through $\alpha_t$. This regime captures non-stationarity without
    exploitable periodic structure and primarily rewards adaptive tracking.

    \item \textbf{Seasonality-driven predictable decline regime.}
    To construct settings where intertemporal allocation decisions matter, we introduce a
    repeatable \emph{week-of-quarter efficiency factor} $g(w)$, where
    $w \in \{1,\dots,12\}$ indexes the week within the quarter. The factor $g(w)$ captures
    systematic within-quarter declines in return efficiency and is defined as
    \begin{equation}
    g(w) = \exp(-\delta w), \qquad \delta > 0 .
    \label{eq:seasonal_decline_exp}
    \end{equation}
    The effective spend--return relationship in week $w$ is then
    \begin{equation}
    \rho_w(s; \theta) = g(w)\,\rho(s; \theta).
    \label{eq:eff_response_exp}
    \end{equation}
    Because this efficiency pattern repeats across quarters, it is identifiable from
    historical data and can be exploited by forward-looking control policies.
\end{enumerate}

%% file: Policies.tex
\section{Budget Allocation: Control Strategies}
\label{sec:control}

We study how a fixed quarterly budget should be allocated across weeks to maximize
cumulative return in the presence of (i) \emph{execution noise}, where planned budgets
do not translate deterministically into realized spend, and (ii) \emph{operational
guardrails}, such as smooth week-to-week budget changes. We compare two policies: a
reactive pacing baseline that relies only on budget accounting, and a receding-horizon
\emph{Model Predictive Control} (MPC) allocator that combines a control-facing forward
model (\emph{prediction module}) with constrained planning (\emph{optimization module}).
Both policies are evaluated under the operating regimes in
Section~\ref{sec:regimes}.

\subsection{Baseline: Reactive Weekly Pacing with Carryover}
\label{sec:baseline}

As a non-predictive reference, we consider a reactive pacing strategy that allocates budget
based solely on remaining funds and time, without modeling or forecasting return
efficiency. This class of strategies is widely used in practice as a feasibility-oriented
approach to budget delivery under execution uncertainty.

The baseline operates at a weekly cadence and consists of two steps: (i) defining nominal
weekly targets using historical pacing ratios, and (ii) updating these targets via
carryover based on realized spend.

\paragraph{Phase 1 (weekly targets from historical ratios).}
Let $B_Q$ denote the total quarterly budget and let $W$ denote the number of weeks in the
quarter. We define a set of historical weekly pacing ratios
$\{\pi_1,\ldots,\pi_W\}$, where $\pi_t \ge 0$ and $\sum_{t=1}^{W} \pi_t = 1$. Each $\pi_t$
represents the fraction of total spend historically allocated to week $t$, computed from
the previous year’s spend pattern over the same quarter (or a comparable reference
period). The nominal weekly budget is defined as
\begin{equation}
B_t^{(0)} = \pi_t\, B_Q, \qquad t = 1,\ldots,W.
\end{equation}
If historical ratios are unavailable, we use a uniform fallback $\pi_t = 1/W$.

\paragraph{Phase 2 (weekly carryover).}
Let $B^{\mathrm{rem}}_t$ denote the remaining budget at the start of week $t$, initialized as
$B^{\mathrm{rem}}_1 = B_Q$. To account for under- or over-delivery in previous weeks, the
planned weekly budget is computed by rescaling the remaining budget according to the
remaining pacing ratios:
\begin{equation}
b_t
=
B^{\mathrm{rem}}_t \cdot
\frac{\pi_t}{\sum_{j=t}^{W} \pi_j}.
\end{equation}
After execution of week $t$, the environment produces realized spend $s_t$, and the
remaining budget is updated via
\begin{equation}
B^{\mathrm{rem}}_{t+1} = B^{\mathrm{rem}}_t - s_t,
\end{equation}
so that any surplus or deficit is automatically carried forward and absorbed into the
remaining weeks.

\textbf{Key property.}
The baseline is purely reactive: weekly budgets depend only on historical pacing ratios,
realized spend, and remaining time, and do not use forecasts of future return efficiency.
It therefore provides a strong benchmark for feasibility and budget utilization, but
cannot exploit predictable temporal structure in returns.

\subsection{Model Predictive Control (MPC): Prediction and Optimization}
\label{sec:mpc}

MPC treats budgeting as a receding-horizon control problem. At each week $\tau$, it
(i) generates horizon forecasts of realized spend and return under candidate budgets,
(ii) solves a constrained finite-horizon planning problem, (iii) executes only the first
weekly budget, and (iv) repeats after observing realized spend and return. We explicitly
separate
\begin{enumerate}
    \item \textbf{Prediction module}: a control-facing forward model producing
    $\widehat{s}_t(\cdot)$ and $\widehat{r}_t(\cdot)$ over the horizon
    \item \textbf{Optimization module}: a constrained planner that maps forecasts into
    implementable weekly budgets under guardrails
\end{enumerate}

\subsubsection{Optimization module (finite-horizon planning)}
\label{sec:mpc_opt}

At week $\tau$, MPC selects a horizon-$H$ budget sequence
$\mathbf{b}_{\tau:\tau+H-1} = \{b_\tau,\ldots,b_{\tau+H-1}\}$ by solving
\begin{align}
\max_{\mathbf{b}_{\tau:\tau+H-1}}\quad
& \sum_{h=0}^{H-1}
\widehat{r}_{\tau+h}\!\Bigl(b_{\tau+h};\widehat{\theta}_{\tau+h|\tau}\Bigr)
\label{eq:mpc_obj_new2} \\
\text{s.t.}\quad
& \sum_{h=0}^{H-1}
\widehat{s}_{\tau+h}\!\bigl(b_{\tau+h}\bigr) \le B^{\mathrm{rem}}_\tau,
\label{eq:mpc_budget_new2} \\
& b_{\tau+h} \ge 0, \qquad h = 0,\ldots,H-1,
\label{eq:mpc_nonneg_new2} \\
& (1-\gamma_L)b_{\tau+h-1} \le b_{\tau+h} \le (1+\gamma_U)b_{\tau+h-1}.
\label{eq:mpc_rel_smooth_new2}
\end{align}

The objective~\eqref{eq:mpc_obj_new2} maximizes predicted cumulative return over the planning
horizon using the control-facing forward model. The budget constraint
\eqref{eq:mpc_budget_new2} enforces feasibility with respect to the remaining quarter budget
$B^{\mathrm{rem}}_\tau = B_Q - \sum_{t<\tau} s_t$ based on realized spend, and is written as
an inequality to accommodate execution frictions. Constraints
\eqref{eq:mpc_nonneg_new2}--\eqref{eq:mpc_rel_smooth_new2} enforce nonnegativity and impose a
relative smoothness guardrail that limits week-to-week changes in planned spend, $\gamma_L$ and $\gamma_U$ provides the lower and upper percentage change bounds. MPC operates in receding horizon by applying the first action
$b_\tau^{\mathrm{mpc}} = b_\tau^\star$, observing realized spend and return $(s_\tau, r_\tau)$,
updating forecasts, and re-solving
\eqref{eq:mpc_obj_new2}--\eqref{eq:mpc_rel_smooth_new2} at $\tau \leftarrow \tau + 1$.

 \textbf{Oracle vs.\ Estimated MPC:} To disentangle the value of receding-horizon planning from the limits of prediction,
we evaluate two MPC variants. We define them as follows:

\begin{enumerate}
    \item \textbf{Oracle MPC:} forms the predicted return model $\widehat{r}_t(\cdot)$ using
    the simulator's true latent state $\theta_t$ and the true generating functional form. This provides an upper-bound on the results.

    \item \textbf{Learned MPC:} forms $\widehat{r}_t(\cdot)$ using parameter estimates
    $\widehat{\theta}_{t|\tau}$ computed from historical data and accrued online observations
    under the regime-specific estimation procedures, together with the assumed functional form
    in~\ref{sec:mpc_pred}.
\end{enumerate}

\subsubsection{Prediction module (nominal nonlinear spend--return model)}
\label{sec:mpc_pred}

Returns are modeled as a nonlinear saturating function of realized spend with latent
parameters $\theta_t$ that may evolve over time. As a nominal model, we use exponential
saturation with $\theta_t = (\rho_{\max,t}, \kappa_t)$, where $\rho_{\max,t} > 0$ is the
saturation return and $\kappa_t > 0$ controls the rate of saturation. For realized spend
$s \ge 0$,
\begin{equation}
\rho(s;\theta_t)
=
\rho_{\max,t}\bigl(1 - e^{-\kappa_t s}\bigr).
\label{eq:exp_sat}
\end{equation}
The return predictor supplied to the optimization module is
\begin{equation}
\widehat{r}_{t}\!\bigl(b;\widehat{\theta}_{t|\tau}\bigr)
=
\rho \Bigl(\widehat{s}_{t}(b);\widehat{\theta}_{t|\tau}\Bigr),
\label{eq:pred_return_theta2}
\end{equation}
which makes the role of predictability explicit: MPC can only outperform reactive pacing
when the forecast sequence
$\{\widehat{\theta}_{\tau+h|\tau}\}_{h=0}^{H-1}$ contains exploitable structure over the
planning horizon.

\paragraph{Implicit saturation spend}
Define the $\eta$-saturation spend $s_\eta$ as the spend required to reach a fraction
$\eta \in (0,1)$ of the saturation return,
$\rho(s_\eta;\rho_{\max,t},\kappa_t) = \eta\,\rho_{\max,t}$, which yields
\begin{equation}
s_\eta(\rho_{\max,t},\kappa_t)
=
-\frac{1}{\kappa_t}\log(1-\eta).
\label{eq:sat_spend_eta}
\end{equation}

\subsubsection{Forecasting $\theta_t$ across operating regimes}
\label{sec:mpc_prediction_regimes2}

The regimes in Section~\ref{sec:regimes} differ only in how the prediction module
constructs the horizon forecasts $\widehat{\theta}_{\tau+h|\tau}$ used in
\eqref{eq:mpc_obj_new2}.

\regimepara{Static}
Estimate a single response curve on historical data and hold parameters fixed within the
quarter
\begin{equation}
\widehat{\theta}_{\tau+h|\tau}
=
\widehat{\theta}_{\tau|\tau}, \qquad h = 0,\ldots,H-1.
\label{eq:theta_static2}
\end{equation}

\regimepara{Random-walk drift regime (PF-MPC)} 
In this regime latent return efficiency evolves stochastically with no exploitable
periodicity. MPC therefore relies on \emph{adaptive tracking rather than forecasting}.
We implement this using a particle-filter-based MPC (PF-MPC), where a latent efficiency
state is updated sequentially under a log-space random-walk assumption and treated as
locally persistent over the planning horizon. The particle filter is updated once per
week using realized spend--return observations, and MPC replans using the current
posterior mean state without assuming future structure.

\regimepara{Forecastable seasonal dynamics} 
When latent parameter trajectories exhibit repeatable structure such as trends or
seasonality, we explicitly forecast them. Time-series predictors are fit to identified
parameter sequences (after suitable transformation, e.g., log space) and extrapolated
over the MPC horizon. For parameters $\theta_t = (\rho_{\max,t}, \kappa_t)$, we generate
forecasts
\begin{align}
\rho_{\max,t+h} &= f_{\rho}\bigl(\{\rho_{\max,\tau}\}_{\tau \le t}, h\bigr), \\
\kappa_{t+h} &= f_{\kappa}\bigl(\{\kappa_{\tau}\}_{\tau \le t}, h\bigr),
\end{align}
for horizons $h \in \{1,\ldots,H\}$. In practice, we employ SARIMAX models to capture both
autoregressive structure and seasonal effects in parameter evolution. When returns
exhibit systematic within-quarter decline, accurate forecasting enables MPC to
anticipate future conditions and allocate budget proactively rather than reactively.
These forecasts are supplied to~\eqref{eq:pred_return_theta2} and optimized via
\eqref{eq:mpc_obj_new2}--\eqref{eq:mpc_rel_smooth_new2}.

%% file: Experimental_Setup.tex
\section{Experimental Setup}
\label{sec:experiments}

All experiments are conducted in the synthetic marketing system defined in
Section~\ref{sec:data_generation}. Unless otherwise stated, all environment dynamics,
noise processes, operating regimes, and budget construction procedures follow that
section exactly. The experimental setup below specifies only the evaluation protocol and
statistical reporting.

\subsection{Evaluation protocol}
\label{sec:evaluation_protocol}

Each trial consists of a historical period used solely for policy initialization,
followed by a held-out evaluation quarter of $W = 12$ weeks. During the evaluation period,
policies interact with the system in closed loop and select weekly budgets based only on
past realized observations. Within each trial, all policies are evaluated under identical
initial conditions and stochastic realizations (paired evaluation).

\subsection{Evaluation metrics}
\label{sec:metrics}

To enable comparison across budgets and operating regimes without reporting absolute
monetary values, we report \emph{mean} returns in oracle-normalized form. Let
$R^{\mathrm{oracle}}$ denote the total return achieved by the oracle allocation under a
given trial realization. For any policy $\pi$, we define the oracle-normalized return
\begin{equation}
\widetilde{R}^{\pi}
=
\frac{R^{\pi}}{R^{\mathrm{oracle}}}.
\end{equation}
By construction, $\widetilde{R}^{\mathrm{oracle}} = 1$. We report
$\mathbb{E}[\widetilde{R}^{\pi}]$ as a scale-free summary of average performance.

In addition, \emph{performance differences} are computed on the original return scale on a
per-trial basis and reported as percentages. Specifically, for each trial we compute:
\begin{align}
\Delta^{\pi}_{\mathrm{base}}
&=
100 \cdot \frac{R^{\pi} - R^{\mathrm{base}}}{R^{\mathrm{base}}},
\\
\mathrm{Gap}^{\pi}_{\mathrm{oracle}}
&=
100 \cdot \frac{R^{\mathrm{oracle}} - R^{\pi}}{R^{\mathrm{oracle}}}.
\end{align}
Confidence intervals for $\Delta^{\pi}_{\mathrm{base}}$ and
$\mathrm{Gap}^{\pi}_{\mathrm{oracle}}$ are obtained from their empirical distributions
across trials using a paired design.

We report the following control-oriented metrics:
\begin{itemize}
    \item \textbf{Normalized return (mean):} $\mathbb{E}[\widetilde{R}^{\pi}]$
    \item \textbf{Budget utilization:} $\sum_{t=1}^{W} s_t \,/\, B^{(q)}$
    \item \textbf{Improvement over baseline (\%):}
    $\Delta^{\pi}_{\mathrm{base}}$ with 95\% confidence intervals
    \item \textbf{Oracle gap (\%):}
    $\mathrm{Gap}^{\pi}_{\mathrm{oracle}}$ with 95\% confidence intervals
\end{itemize}

\subsection{Monte Carlo evaluation}
\label{sec:monte_carlo}

Each configuration is evaluated over $N$ independent trials. For each trial, all policies
share the same historical data and the same evaluation realization. We report the mean and
95\% confidence intervals for $\Delta$ using both a paired $t$-interval and a nonparametric
bootstrap. Statistical significance is assessed using a paired $t$-test on per-trial
return differences.

%% file: Experimental_Results.tex
\section{Experimental Results}

We evaluate reactive budgeting and receding-horizon Model Predictive Control (MPC) under
increasingly structured forms of non-stationarity. Our objective is not to compare
estimation accuracy in isolation, but to determine \emph{when predictive control
materially improves closed-loop allocation performance}. All results are computed using
paired Monte Carlo evaluation under identical budgets, constraints, and stochastic
realizations.

We organize the analysis around three research questions:
\begin{enumerate}[label=\textbf{RQ\arabic*.}]
\item \textit{In a \emph{static, time-invariant} response regime, do reactive and predictive
controllers achieve equivalent performance under identical constraints?}
\item \textit{When latent response parameters evolve as a stochastic random walk without
forecastable structure, can predictive control outperform reactive adaptation?}
\item \textit{When latent parameters exhibit learnable seasonal or trend structure, under what
conditions does MPC exploit forecasts to reduce regret relative to myopic policies?}
\end{enumerate}

\subsection{RQ1: Static Regime Sanity Check}
\label{sec:results_static}

\textbf{Key finding:} \textit{MPC provides no meaningful advantage in static environments.}
In stable, time-invariant spend--return regimes with no exploitable temporal structure,
reactive pacing heuristics and receding-horizon MPC converge to equivalent uniform
allocations under identical constraints. Because there is no predictability to exploit,
MPC’s forecasting and lookahead capabilities do not translate into improved closed-loop
performance. Detailed results are reported in Appendix~\ref{sec:rq1_static}.

\subsection{RQ2: Stochastic Drift Without Predictability}
\label{sec:results_drift}

We next examine regimes in which return efficiency evolves over time but lacks exploitable
temporal structure. Latent response parameters follow a stochastic random walk in log space,
\begin{equation}
\log \theta_{t+1}
=
\log \theta_t + \eta_t,
\qquad
\eta_t \sim \mathcal{N}(0,\sigma_{\mathrm{rw}}^2),
\end{equation}
producing smooth yet intrinsically unpredictable dynamics beyond the current state. This
regime isolates \emph{adaptation} from \emph{anticipation}: parameters drift, but there is
no repeatable structure that a predictive controller can reliably exploit over the
planning horizon.

We compare a particle-filter-based MPC (PF-MPC) against the reactive baseline under two
levels of drift intensity: \textbf{mild}
($\sigma_{\mathrm{rw}} \in [0.02, 0.03]$) and
\textbf{moderate} ($\sigma_{\mathrm{rw}} \in [0.05, 0.08]$). Mean returns are reported in
oracle-normalized form to remove scale effects, while performance gaps and confidence
intervals are computed on the original per-trial return scale and reported as percentages.

\textbf{Key findings:}
\begin{itemize}
    \item \textbf{No gain from prediction without predictability.}
    PF-MPC underperforms the reactive baseline under both drift intensities, with mean
    performance gaps on the order of $0.3$--$0.4\%$, despite both policies fully exhausting
    the available budget.

    \item \textbf{Limits of adaptive tracking.}
    While PF-MPC adapts to local parameter changes, model misspecification and tracking
    error lead to conservative allocations that forgo upside relative to reactive pacing.

    \item \textbf{Downside mitigation in difficult realizations.}
    PF-MPC outperforms the baseline in a nontrivial fraction of trials, primarily when
    baseline performance is poor. This indicates that adaptivity can reduce downside risk
    even when it does not deliver consistent average gains.

    \item \textbf{Oracle gap reflects missing predictability.}
    The oracle gap increases with drift intensity, revealing substantial theoretical
    headroom that cannot be realized in the absence of forecastable temporal structure.
\end{itemize}

\begin{table}[t]
\centering
\footnotesize
\setlength{\tabcolsep}{4pt}
\renewcommand{\arraystretch}{0.95}
\caption{Performance under stochastic random-walk drift. Mean returns are
reported relative to the oracle allocation. Percentage improvements and oracle gaps are
computed on the original return scale.}
\label{tab:dynamic_results}
\begin{tabular}{lcccc}
\toprule
\textbf{Drift Strength} &
\textbf{Baseline} &
\textbf{PF-MPC} &
\textbf{PF-MPC Improve.} &
\textbf{Oracle Gap} \\
\midrule
Mild
& 0.696
& 0.694
& $-0.38\%$
& $+45.2\%$ \\
& 
& 
& $[-0.48, -0.28]\%$
& $[41.9, 48.5]\%$ \\
Moderate
& 0.678
& 0.675
& $-0.31\%$
& $+51.9\%$ \\
& 
& 
& $[-0.53, -0.08]\%$
& $[46.8, 57.1]\%$ \\
\bottomrule
\end{tabular}
\end{table}

\textbf{Interpretation:}
In stochastic drift regimes, efficiency fluctuations average out over the planning horizon,
leaving little structure for anticipatory control. As a result, receding-horizon MPC
effectively collapses to a re-optimized reactive policy. Time variation alone is therefore
insufficient for long-horizon gains; predictive control requires predictable structure to
deliver systematic improvements.

\subsection{RQ3: Predictable Seasonal Dynamics}
\label{sec:results_seasonal}

We finally consider regimes in which return efficiency exhibits a repeatable
within-quarter decline, capturing effects such as ad fatigue or lifecycle decay. Let
$L_t$ denote the multiplicative return-efficiency level in week $t$, with $L_0>0$
representing the initial efficiency at the start of the quarter. Efficiency evolves
deterministically according to
\begin{equation}
L_t = L_0 (1-\delta)^t,
\end{equation}
where $\delta \in \{0.05, 0.10, 0.15, 0.20\}$ controls the rate of within-quarter decline,
spanning weak to strong seasonal structure. Because this pattern repeats across quarters,
it is identifiable from historical data and forecastable by forward-looking control
policies.

Mean total returns are reported in oracle-normalized form for scale-free comparison.
Performance improvements relative to the baseline and oracle gaps are computed on the
original per-trial return scale and reported as percentages with 95\% confidence
intervals.

\textbf{Key findings:}
\begin{itemize}
    \item \textbf{Threshold behavior.}
    MPC underperforms reactive pacing under weak seasonality ($\delta = 5\%$), but
    consistently outperforms the baseline once the decline exceeds approximately
    $\delta \approx 10\%$.

    \item \textbf{Scaling with predictability.}
    Performance gains increase sharply with the strength of seasonal structure, indicating
    that intertemporal reallocation becomes increasingly valuable as predictability
    strengthens.

    \item \textbf{Intertemporal allocation mechanism.}
    MPC shifts spend toward earlier, higher-efficiency weeks, while the baseline spreads
    spend uniformly and incurs increasing opportunity cost as decline intensifies.

    \item \textbf{High efficiency relative to the oracle.}
    Across all decline rates, MPC captures the majority of the achievable gains. The oracle
    gap remains stable and modest, indicating that operational smoothness constraints—not
    forecasting limitations—are the dominant bottleneck.
\end{itemize}

\begin{table}[t]
\centering
\footnotesize
\setlength{\tabcolsep}{3pt}
\renewcommand{\arraystretch}{0.95}
\caption{Performance under predictable seasonal decline. Mean returns are
reported relative to the oracle allocation. Percentage improvements and oracle gaps are
computed on the original return scale.}
\label{tab:seasonal_results}
\begin{tabular}{lccccc}
\hline
\textbf{$\delta$} &
\textbf{Baseline} &
\textbf{MPC} &
\textbf{MPC Improve.} &
\textbf{95\% CI} &
\textbf{Oracle Gap} \\
\hline
5\%  & 0.972 & 0.939 & $-3.2\%$  & $[-3.53, -2.92]\%$ & $+6.0\%$ \\
10\% & 0.905 & 0.933 & $+4.2\%$  & $[3.98, 4.38]\%$   & $+6.9\%$ \\
15\% & 0.852 & 0.931 & $+11.2\%$ & $[10.68, 11.72]\%$ & $+7.2\%$ \\
20\% & 0.808 & 0.944 & $+14.0\%$ & $[13.13, 14.89]\%$ & $+6.8\%$ \\
\hline
\end{tabular}
\end{table}

\paragraph{Interpretation.}
Predictable seasonal decline provides a stable forecast signal that enables effective
long-horizon planning. MPC exploits this structure by reallocating spend toward earlier,
higher-efficiency periods, yielding increasing gains as seasonality strengthens. The
stable and modest oracle gap across all decline rates indicates that under the imposed
$\pm 20\%$ smoothness constraints, MPC already captures most of the theoretically
achievable benefit from predictability. Further gains would therefore require relaxing
operational constraints rather than improving forecasting accuracy.

\textbf{Interpretation:}
Predictable seasonal decline provides a stable forecast signal that enables effective
long-horizon planning. MPC exploits this structure by reallocating spend toward earlier,
higher-efficiency periods, yielding increasing gains as the strength of seasonality
increases. The consistently modest oracle gap across all decline rates, together with
tight confidence intervals, indicates that under the imposed $\pm 20\%$ smoothness
constraints MPC already captures most of the theoretically achievable benefit from
predictability. Further gains would therefore require relaxing operational constraints
rather than improving forecasting accuracy.

%% file: Conclusion.tex
\section{Conclusion}
This work studies dynamic system identification for marketing spend--return processes, emphasizing the role of identifiable temporal structure in enabling predictive optimization rather than fitting static response curves. Through simulation under execution noise, parameter drift, and model misspecification, we show that state-space identification can approximate the data-generating process and narrow the gap between realized and oracle behavior, with MPC gains emerging only when predictable dynamics are present. The experimental scope is intentionally constrained, focusing on a controlled form of seasonality and simplified assumptions such as homoskedastic Gaussian noise and limited latent-state structure. Extending this framework to richer identification choices—including alternative state parameterizations, non-Gaussian and heteroskedastic noise, regime shifts, and more complex seasonal patterns—remains an important direction for future work and is necessary to fully characterize when improved identification fidelity translates into actionable control value in realistic budget planning settings.

%% file: Appendix.tex
\appendices

\section{Implementation Details}
\label{sec:mpc_impl}

The optimization problem
\eqref{eq:mpc_obj_new2}--\eqref{eq:mpc_rel_smooth_new2}
is a small nonlinear program solved on a weekly cadence using standard solvers
(e.g., CasADi/IPOPT). All results report end-to-end closed-loop performance.
Forecasting components are not tuned in isolation; identification quality is
evaluated solely through its downstream impact on control outcomes.

\section{Budget Construction Details}
\label{app:budget_construction}

Let $\{s_d\}_{d=1}^{D}$ denote daily realized spend in the final year of the
historical period, where $D$ is the number of days in that year under the
simulation discretization. Annual spend is computed as
\begin{equation}
S_{\mathrm{year}} = \sum_{d=1}^{D} s_d .
\end{equation}

To introduce realistic year-over-year variation without incorporating any
information from the evaluation period, we apply a multiplicative scaling
factor $\lambda \sim \mathcal{D}_\lambda$, yielding a scaled annual budget
\begin{equation}
\widetilde{S}_{\mathrm{year}} = \lambda \, S_{\mathrm{year}} .
\end{equation}

Monthly allocation proportions $\{\rho_m\}_{m=1}^{12}$ are estimated from
historical data as
\begin{equation}
\rho_m = \frac{H_m}{\sum_{j=1}^{12} H_j},
\end{equation}
where $H_m$ denotes total spend in month $m$ of the historical year. The budget
assigned to month $m$ is then
\begin{equation}
B_m = \rho_m \, \widetilde{S}_{\mathrm{year}} .
\end{equation}

For an evaluation quarter $q \subset \{1,\dots,12\}$ consisting of three
consecutive months, the quarterly budget is defined as
\begin{equation}
B^{(q)} = \sum_{m \in q} B_m .
\end{equation}

All allocation policies are initialized using historical data only and make
decisions during the evaluation quarter based exclusively on past realized
spend and return observations, ensuring a strict separation between
initialization and evaluation.

\section{RQ1 Details: Static (Stationary) Environment}
\label{sec:rq1_static}

We first consider a static regime in which the spend--return relationship is time-invariant
over the evaluation horizon. In this setting, response parameters can be consistently
estimated from historical data, and under concave returns the optimal feasible allocation
is approximately uniform across time, up to budget feasibility and smoothness constraints.
In the absence of predictable temporal variation, receding-horizon planning offers no
strategic advantage over reactive pacing.

Consistent with this expectation, MPC and the baseline achieve nearly identical performance
when evaluated relative to the oracle benchmark. Table~\ref{tab:static_results} reports
oracle-normalized mean returns, defined as the ratio of cumulative return achieved by a
policy to that of the oracle allocation under identical realizations. By construction, the
oracle attains a normalized return of one. Both MPC and the baseline lie within
approximately $0.5\%$ of this upper bound.

Performance differences are small and statistically significant only due to constraint
friction and forecast noise rather than anticipatory reallocation. The oracle gap is
negligible, confirming that even perfect foresight provides little benefit in static
environments. Both MPC and the oracle operate under identical quarterly budget constraints
and $\pm 30\%$ relative smoothness limits, which restrict aggressive back-loading even when
the true response function is known. As a result, lookahead optimization does not improve
outcomes over reactive pacing when no exploitable temporal structure is present.

\begin{table}[h]
\centering
\setlength{\tabcolsep}{4pt}
\renewcommand{\arraystretch}{0.95}
\caption{Allocator performance in the static regime. Mean returns are
reported relative to the oracle allocation. Percentage improvements and oracle gaps are
computed on the original return scale.}
\label{tab:static_results}
\begin{tabular}{lcccc}
\hline
\textbf{Strategy} &
\textbf{Normalized Return} &
\textbf{95\% CI} &
\textbf{Improvement} &
\textbf{Oracle Gap} \\
\hline
Baseline
& 1.000
& $[0.986, 1.014]$
& --- 
& $+0.00\%$ \\
MPC
& 0.997
& $[0.981, 1.011]$
& $-0.37\%$
& $+0.34\%$ \\
\hline
\end{tabular}
\end{table}

%% file: bib.bib
@inproceedings{chen2011real,
  title={Real-time bidding algorithms for performance-based display ad allocation},
  author={Chen, Ye and Berkhin, Pavel and Anderson, Bo and Devanur, Nikhil R},
  booktitle={Proceedings of the 17th ACM SIGKDD international conference on Knowledge discovery and data mining},
  year={2011}
}

@inproceedings{Wang2023,
  author    = {Jing Wang and Chris Swartz and Kai Huang and Smriti Shyamal and Roopesh Ranjan and Dan Friedman and Joel Brooks},
  title     = {Data-driven Budget Allocation Optimization for Digital Marketing},
  booktitle = {Proceedings of the 65th Annual Canadian Operational Research Society Conference},
  address   = {London, Ontario},
  year      = {2024}
}

@book{hanssens2003market,
  title={Market Response Models: Econometric and Time Series Analysis},
  author={Hanssens, Dominique M. and Parsons, Leonard J. and Schultz, Randall L.},
  year={2003},
  publisher={Springer}
}

@article{luzon2022dynamic,
  title={Dynamic budget allocation in social media advertising campaigns},
  author={Luzon, Virginia},
  journal={European Journal of Operational Research},
  volume={298},
  number={1},
  pages={327--341},
  year={2022},
  publisher={Elsevier},
  doi={10.1016/j.ejor.2021.07.024}
}

@article{yan2023endtoend,
  title={An end-to-end framework for marketing effectiveness optimization under budget constraint},
  author={Yan, Ziang and Wang, Shusen and Zhou, Guorui and Lin, Jingjian and Jiang, Peng},
  journal={arXiv preprint arXiv:2302.04477},
  year={2023}
}

@inproceedings{hatano2015lagrangian,
  title={Lagrangian decomposition algorithm for allocating marketing channels},
  author={Hatano, Daisuke and Fukunaga, Takuro and Maehara, Takanori and Kawarabayashi, Ken-ichi},
  booktitle={Proceedings of the AAAI Conference on Artificial Intelligence},
  year={2015}
}

@inproceedings{archak2012budget,
  title={Budget optimization for online campaigns with positive carryover effects},
  author={Archak, Nikolay and Mirrokni, Vahab and Muthukrishnan, Shanmugavelayutham},
  booktitle={International Workshop on Internet and Network Economics},
  pages={86--99},
  year={2012},
  organization={Springer}
}

@article{nerlove1962optimal,
  title={Optimal Advertising Policy Under Dynamic Conditions},
  author={Nerlove, Marc and Arrow, Kenneth J.},
  journal={Economica},
  volume={29},
  number={114},
  pages={129--142},
  year={1962},
  publisher={Wiley}
}

@article{sethi1973optimal,
  title={Optimal Control of the Vidale--Wolfe Advertising Model},
  author={Sethi, Suresh P.},
  journal={Operations Research},
  volume={21},
  number={4},
  pages={998--1013},
  year={1973},
  publisher={INFORMS}
}

@book{morari1999model,
  title={Model predictive control: past, present and future},
  author={Morari, Manfred and Lee, Jay H.},
  journal={Computers \& Chemical Engineering},
  pages={667--682},
  year={1999},
  publisher={Elsevier}
}

@article{DekimpeHanssens1995Persistence,
  author  = {Dekimpe, Marnik G. and Hanssens, Dominique M.},
  title   = {The Persistence of Marketing Effects on Sales},
  journal = {Marketing Science},
  year    = {1995},
  volume  = {14},
  number  = {1},
  pages   = {1--21},
  doi     = {10.1287/mksc.14.1.1}
}

@inproceedings{zhao2019unified,
  title={A Unified Framework for Advertising Bidding and Budget Allocation},
  author={Zhao, Qingpeng and Liu, Yu and Wei, Wenyu and Chen, Wei and Wang, Jun},
  booktitle={Proceedings of the 25th ACM SIGKDD International Conference on Knowledge Discovery \& Data Mining},
  year={2019},
  pages={2185--2194}
}

@inproceedings{cai2017realtime,
  title={Real-Time Bidding by Reinforcement Learning in Display Advertising},
  author={Cai, Hongyi and Ren, Kan and Zhang, Weinan and Malialis, Kleanthis and Wang, Jun and Yu, Yong and Guo, Dawei},
  booktitle={Proceedings of the Tenth ACM International Conference on Web Search and Data Mining},
  year={2017},
  pages={661--670}
}

@inproceedings{cai2023kdd_autobid,
  title={Budget-Constrained Marketing Optimization via Automated Bidding},
  author={Cai, Hongyi and Chen, Wei and Wang, Jun},
  booktitle={Proceedings of the 29th ACM SIGKDD Conference on Knowledge Discovery and Data Mining},
  year={2023}
}

@book{rawlings2017model,
  title={Model Predictive Control: Theory, Computation, and Design},
  author={Rawlings, James B. and Mayne, David Q. and Diehl, Moritz},
  year={2017},
  publisher={Nob Hill Publishing},
  edition={2nd}
}

@inproceedings{agarwal2014budgetpacing,
  title={Budget Pacing for Targeted Online Advertisements at LinkedIn},
  author={Agarwal, Deepak and others},
  booktitle={Proceedings of the 20th ACM SIGKDD international conference on Knowledge discovery and data mining},
  pages={1613--1619},
  year={2014}
}

@article{balseiro2024fieldguide,
  title={A field guide for pacing budget and ros constraints},
  author={Balseiro, Santiago R and Bhawalkar, Kshipra and Feng, Zhe and Lu, Haihao and Mirrokni, Vahab and Sivan, Balasubramanian and Wang, Di},
  journal={arXiv preprint arXiv:2302.08530},
  year={2023}
}

@misc{expedia10k,
  title        = {Expedia Group, Inc. Form 10-K Annual Report},
  author       = {{Expedia Group, Inc.}},
  year         = {2024},
  howpublished = {\url{https://www.sec.gov/ixviewer/documents/20240208/0001637459-24-000014.xhtml}},
  note         = {Accessed via U.S. SEC EDGAR}
}

@misc{expedia10q,
  title        = {Expedia Group, Inc. Form 10-Q Quarterly Reports},
  author       = {{Expedia Group, Inc.}},
  year         = {2024},
  howpublished = {\url{https://www.sec.gov/cgi-bin/browse-edgar?action=getcompany&CIK=0001637459&type=10-Q}},
  note         = {Accessed via U.S. SEC EDGAR}
}

@article{li2019online,
  title={Online optimal control with linear dynamics and predictions: Algorithms and regret analysis},
  author={Li, Yingying and Chen, Xin and Li, Na},
  journal={Advances in Neural Information Processing Systems},
  volume={32},
  year={2019}
}

@article{yu2020power,
  title={The power of predictions in online control},
  author={Yu, Chenkai and Shi, Guanya and Chung, Soon-Jo and Yue, Yisong and Wierman, Adam},
  journal={Advances in Neural Information Processing Systems},
  volume={33},
  pages={1994--2004},
  year={2020}
}

@article{richards1959flexible,
  title={A flexible growth function for empirical use},
  author={Richards, Francis J},
  journal={Journal of experimental Botany},
  volume={10},
  number={2},
  pages={290--301},
  year={1959},
  publisher={Oxford University Press}
}
